%% file: dpf2013talk_final.tex
\documentclass[12pt]{article}
\usepackage{graphicx}
\DeclareGraphicsExtensions{.eps,.ps,.eps.gz,.ps.gz,.png,.jpg}
\usepackage{float, subfigure}
\usepackage{comment}

\newcommand\pubnumber{DPF2013-169, 274}
\newcommand\pubdate{\today}

\def\napoli{$^{1}$Department of Physics, Panjab University, Chandigarh - 160014, India \\
$^{2}$Fermi National Accelerator Laboratory, Batavia, Illinois - 60510, USA}

\def\Title#1{\begin{center} {\Large #1 } \end{center}}
\def\Author#1{\begin{center}{ #1} \end{center}}
\def\Address#1{\begin{center}{ \it #1} \end{center}}

\newcommand\pubblock{\rightline{\begin{tabular}{l} \pubnumber\\
         \pubdate  \end{tabular}}}
\newenvironment{Abstract}{\begin{quotation}  }{\end{quotation}}
\newenvironment{Presented}{\begin{quotation} \begin{center} 
             PRESENTED AT\end{center}\bigskip 
      \begin{center}\begin{large}}{\end{large}\end{center} \end{quotation}}

\input econfmacros.tex
\begin{document}
\begin{titlepage}
\pubblock
\vfill
\Title{Particle Production Measurements using the MIPP Detector at Fermilab}
\vfill
\Author{\textsc{Sonam Mahajan$^{1}$} and \textsc{Rajendran Raja$^{2}$}}
\Address{\napoli}
\begin{center}
\large
For the MIPP Collaboration
\end{center}
\vfill
\begin{Abstract}
The Main Injector Particle Production (MIPP) experiment is a fixed target hadron production experiment at Fermilab. It measures particle production in interactions of 120 GeV/c primary protons from the Main Injector and secondary beams of $\pi^{\pm}, \rm{K}^{\pm}$, p and $\bar{\rm{p}}$ from 5 to 90 GeV/c on nuclear targets which include H, Be, C, Bi and U, and a dedicated run with the NuMI target. MIPP is a high acceptance spectrometer which provides excellent charged particle identification using Time Projection Chamber (TPC), Time of Flight (ToF), multicell Cherenkov (CKOV), Ring Imaging Cherenkov (RICH) detectors, and Calorimeter for neutrons. We present inelastic cross section measurements for 58 and 85 GeV/c p-H interactions, and 58 and 120 GeV/c p-C interactions. A new method is described to account for the low multiplicity inefficiencies in the interaction trigger using KNO scaling. Inelastic cross sections as a function of multiplicity are also presented. The MIPP data are compared with the Monte Carlo predictions and previous measurements. We also describe an algorithm to identify charged particles ($\pi^{\pm}/\rm{p}/\bar{\rm{p}}$ etc.), and present the charged pion and kaon spectra from the interactions of 120 GeV/c protons with carbon target.
\end{Abstract}
\vfill
\begin{Presented}
DPF 2013\\
The Meeting of the American Physical Society\\
Division of Particles and Fields\\
Santa Cruz, California, August 13--17, 2013\\
\end{Presented}
\vfill
\end{titlepage}
\def\thefootnote{\fnsymbol{footnote}}
\setcounter{footnote}{0}
\section{Introduction}
The Main Injector Particle Production (MIPP) experiment is a fixed target hadron production experiment at Fermilab which was operated from December 2004 to February 2006. It used a primary beam of 120 GeV/c protons from the Main Injector and these protons impinged on a copper target to produce secondary beams of charged pions, kaons, protons and anti-protons from 5 to 90 GeV/c~\cite{proc}. The experiment was designed to measure the total charged particle production of $\pi^{\pm}$, K$^{\pm}$, p and $\bar{p}$ off various nuclei including liquid hydrogen (LH$_{2}$), NuMI target and thin targets of beryllium, carbon, bismuth and uranium. \\
The MIPP is a high acceptance spectrometer and it provides excellent charged particle identification using Time Projection Chamber (TPC), Time of Flight (ToF), multicell Cherenkov (CKOV), Ring Imaging Cherenkov (RICH) detectors, and Calorimeter for neutrons. The lay-out of the experiment is shown in Figure 1. 
\section{Motivation}
Monte Carlo (MC) simulation programs such as Geant4, MARS, Fluka, etc. model hadronic interactions based on available data. MIPP is a high acceptance spectrometer and has high statistics data with 6 beam species. These data could be used to improve the hadronic shower simulations.
\section{Inelastic cross section measurements}
Inelastic cross section measurements have been done for 58 and 85 GeV/c proton interactions with LH$_{2}$ target, and 58 and 120 GeV/c proton interactions with carbon target. The LH$_{2}$ target is 14 cm long having 1.5\% interaction length and carbon target is in the form of a disc having 1 cm thickness and 2 inch diameter, and 2\% interaction length. Good events are selected using a cut on transverse positions of incident beam to select good beam tracks, primary vertex selection, rejection of elastics (if any) etc. The interactions are selected using interacton trigger. MIPP used a scintillator-based interaction trigger which requires at least 3 charged tracks for the scintillator to fire. This causes inefficiencies at the low multiplicities and these have to be corrected for. The formula used for cross section calculation is given below:
\begin{figure}
\begin{center}
\includegraphics[width=10cm,height=6cm]{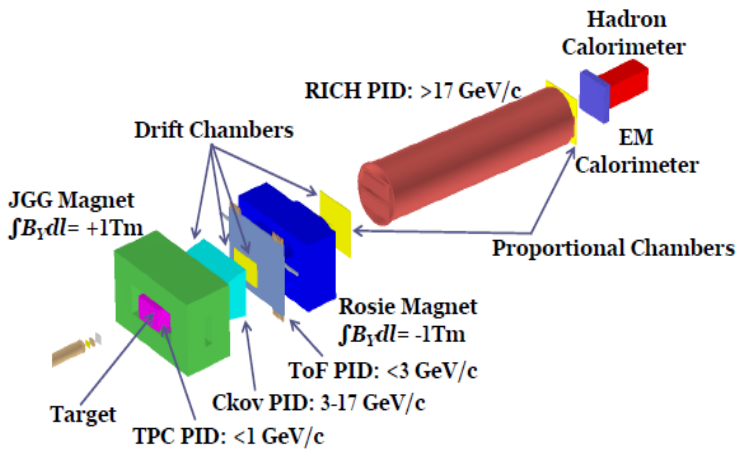}
\caption{The lay-out of the MIPP spectrometer.}
\end{center}
\end{figure}
\begin{center}
$\sigma$ = $\frac{N_{int} \times 10000}{N_{beam} \times n_{t} \times \epsilon}$ mb,  n$_{t}$ = $\frac{N_{A} \times density \times thickness}{Atomic weight}$ cm$^{-2}$
\end{center}
where N$_{int}$ is the number of interactions, N$_{beam}$ is the number of beam particles, n$_{t}$ is the number of target particles per cm$^{2}$ and $\epsilon$ is the product of efficiencies which include the trigger efficiency, cut efficiency and acceptance. The MIPP data are compared with the DPMJET and FLUKA Monte Carlos, and previous measurements for LH$_{2}$ and carbon target in Tables 1 and 2 respectively. The numbers given here are after applying all the corrections which are calculated from the MC. Both the statistical and systematic uncertainties are given. The statistical uncertainty is calculated based on the number of events, and for the systematic uncertainty, we have taken into account the contributions from the beam flux and corrections applied. The systematics from the beam flux is the dominated one, and is $\sim$5\%.\\
\begin{table}[t]
\small
\begin{center}
\scalebox{0.9}{
\begin{tabular}{l|ccc}  
Energy (GeV) & PDG (mb) & DPMJET (mb) & MIPP (mb)  \\ \hline
 58 &   31.13$\pm0.42\rm(stat+syst)$   &  31.6  &  30.33$\pm0.92\rm(stat)^{+2.39}_{-2.48}$(syst)  \\
 85 &   31.42$\pm0.52\rm(stat+syst)$  &  31.8  &  30.65$\pm0.53\rm(stat)^{+2.42}_{-2.52}$(syst)  \\ \hline
\end{tabular}}
\caption{Comparison of the MIPP data with the MC predictions and PDG for p-H interactions at 58 and 85 GeV.}
\end{center}
\end{table}
\begin{table}[t]
\small
\begin{center}
\scalebox{0.9}{
\begin{tabular}{l|ccc}  
Energy (GeV) & Previous measurements (mb) & FLUKA (mb) & MIPP (mb)  \\ \hline
 58 &   252$\pm4.73\rm(stat+syst)$~\cite{publ1}  &  239.1  &  267.6$\pm12.5\rm(stat)^{+17.6}_{-18.6}$(syst)  \\
    &   222$\pm7\rm(stat+syst)$~\cite{publ2}   &     & \\
 120 &        &  240.2  &  264.6$\pm2.94\rm(stat)^{+17.5}_{-18.5}$(syst) \\ \hline
\end{tabular}}
\caption{Comparison of the MIPP data with the MC predictions and previous measurements for p-C interactions at 58 and 120 GeV.}
\end{center}
\end{table}
The comparison is also shown in the form of graph in Figures 2a and 2b for LH$_{2}$ and carbon target respectively. The hydrogen data at both energies are consistent, within error bars, with the PDG and DPMJET. The carbon data at 58 GeV is consistent, within error bars, with the measurement of S. P. Denisov et al.~\cite{publ1} and $\sim$20\% higher than the measurement of A. S. Carroll et al.~\cite{publ2}. FLUKA is $\sim$11\% lower than the 58 GeV/c p-C data. There is no previous measurement for 120 GeV/c p-C interactions. FLUKA is $\sim$9\% lower than the 120 GeV/c p-C data.
\begin{figure*}[ht]
\begin{center}
\subfigure[]{
\includegraphics[width=2.5in,height=2.3in]{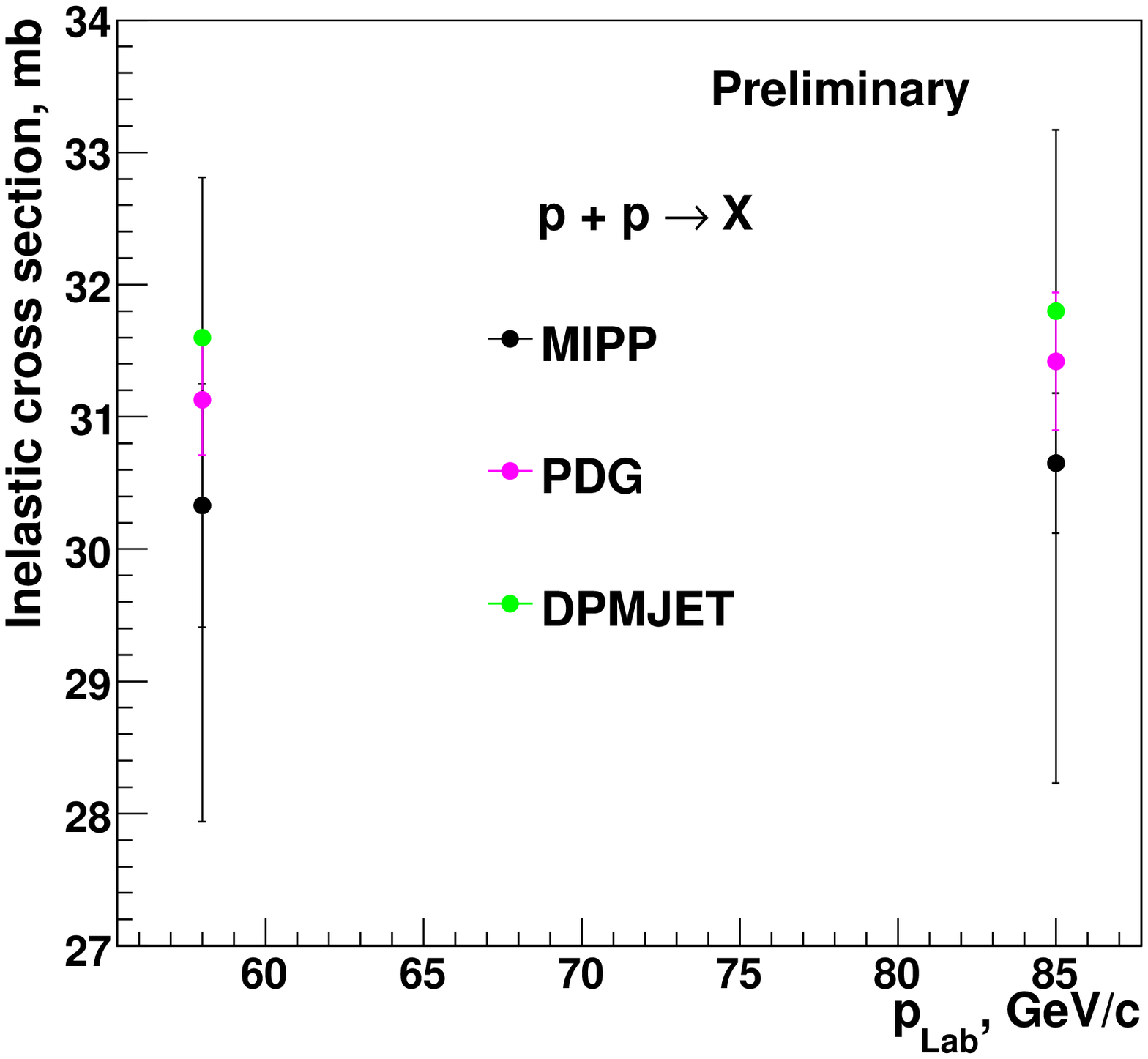}
}
\subfigure[]{
\includegraphics[width=2.5in,height=2.3in]{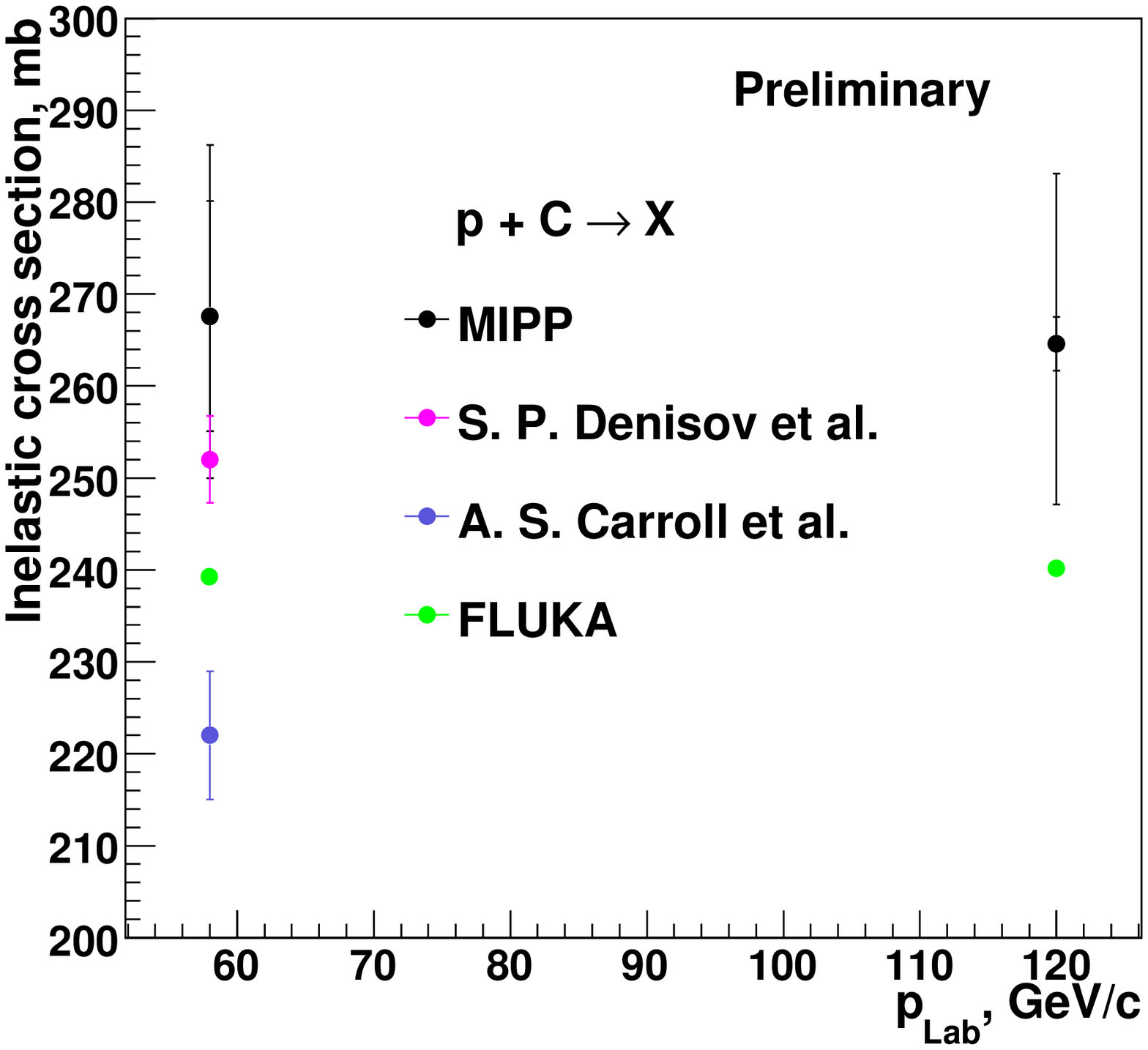}
}
\caption{Comparison of the MIPP data with the MC predictions and previous measurements for p-H (a) and p-C (b) interactions. Both statistical and systematic uncertainties are shown.}
\end{center}
\end{figure*}
\section{KNO-based technique to get trigger efficiency}
A KNO-based technique has been developed to calculate the trigger efficiency. KNO stands for Koba, Nielsen and Olesen, the three authors who put forward the hypothesis that the probability distributions P$_{n}$(s) of producing n charged particles in a certain collision process should exhibit the scaling relation 
\begin{center}
P$_{n}$(s) = $\frac{1}{<n(s)>}\psi(\frac{n}{<n(s)>})$, $\frac{n}{<n(s)>}$ = Z
\end{center}
as s $\rightarrow \infty$ with $<$n(s)$>$ being the average multiplicity of secondaries at collision energy s. $\psi$(Z) is called the KNO scaling function. The KNO scaling states that the function $\psi$(Z) is independent of the energy and only dependent on the variable Z. \\
In this method, we use a K-matrix K(n$_{o}|n_{t}$) from the MC which denotes the probability of obtaining observed multiplicity n$_{o}$, given a true multiplicity n$_{t}$. The trigger is not required in the formation of this matrix. The K-matrix is multiplied by true probabilities from the KNO function~\cite{kno} to get the predicted distribution. The observed distribution where the trigger is applied, is then fitted to the predicted distribution to extract the trigger efficiencies. The trigger efficiencies are the parameters fitted here. \\
The comparison of the observed and predicted distributions at the minimum and the comparison of KNO-based and MC trigger efficiencies as a function of number of tracks passing through the scintillator for 58 GeV/c p-H interactions are shown in Figures 3a and 3b respectively. The observed and predicted distributions agree very well. A drop in the trigger efficiency is observed at n=4 and n=6 which is not reasonable as the trigger efficiency should monotonically increase with the number of tracks through the scintillator. From the studies done, it has been found that this drop occurs due to the discrepancies between the MC and data multiplicity shapes. We have crosschecked the inelastic cross sections using this method and they are within 10\%.
\begin{figure*}[ht]
\begin{center}
\subfigure[]{
\includegraphics[width=2.5in,height=2.3in]{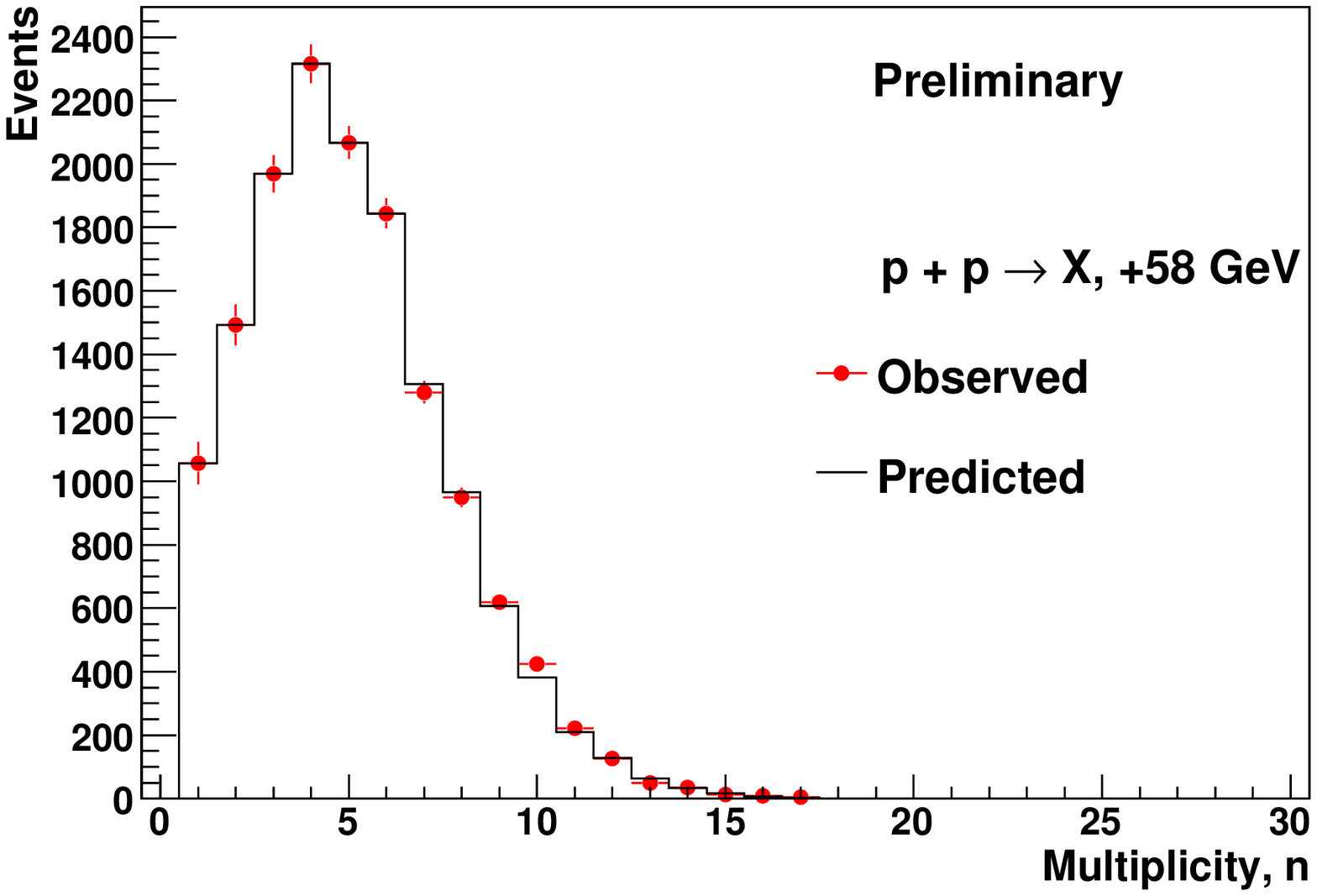}
}
\subfigure[]{
\includegraphics[width=2.5in,height=2.3in]{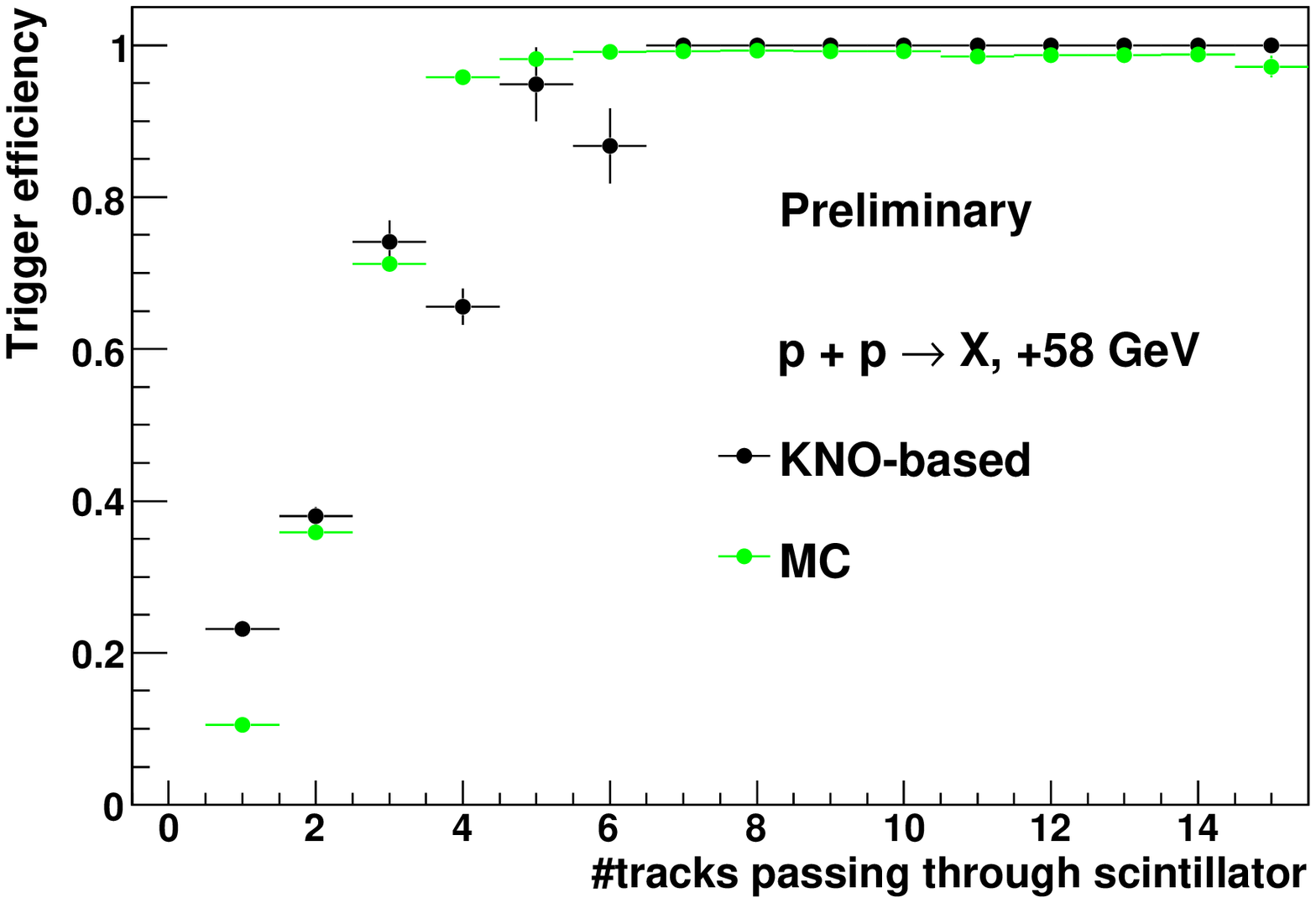}
}
\caption{(a) Comparison of observed and predicted distribution at the minimum. (b) Comparison of KNO-based and MC trigger efficiencies as a function of number of tracks passing through the scintillator for 58 GeV/c p-H interactions.}
\end{center}
\end{figure*}
\section{LH$_{2}$ and carbon multiplicities}
We know that charged multiplicities should be even in case of hydrogen target. Both even and odd multiplicities are observed in the data because of acceptances and reconstruction inefficiencies. We can unfold the MC K-matrix to go back to the data truth. As discrepancies have been observed between the data and MC multiplicity shapes, it has now been decided to use the KNO scaling function to get the data true multiplicity probabilities. The $<$n$>$ from our data is used. Probabilities are multiplied by the average inelastic cross section to get the cross sections as a function of multiplicity. Similarly cross sections are calculated for carbon target where multiplicities are both odd and even.\\
The comparison of the MIPP data with the MC predictions and previous measurements for 58 GeV/c p-H and p-C interactions is shown in Figures 4a and 4b respectively. For LH$_{2}$ target, discrepancies are found between the data and PDG at the lower end and tails. For carbon target, the data is consistent, within error bars, with the measurement of S. P. Denisov et al.~\cite{publ2} and consistent with the measurement of A. S. Carroll et al.~\cite{publ2} for multiplicity$>$15 only. The DPMJET and FLUKA shapes are not consistent with the data.
\begin{figure*}[ht]
\begin{center}
\subfigure[]{
\includegraphics[width=2.5in,height=2.3in]{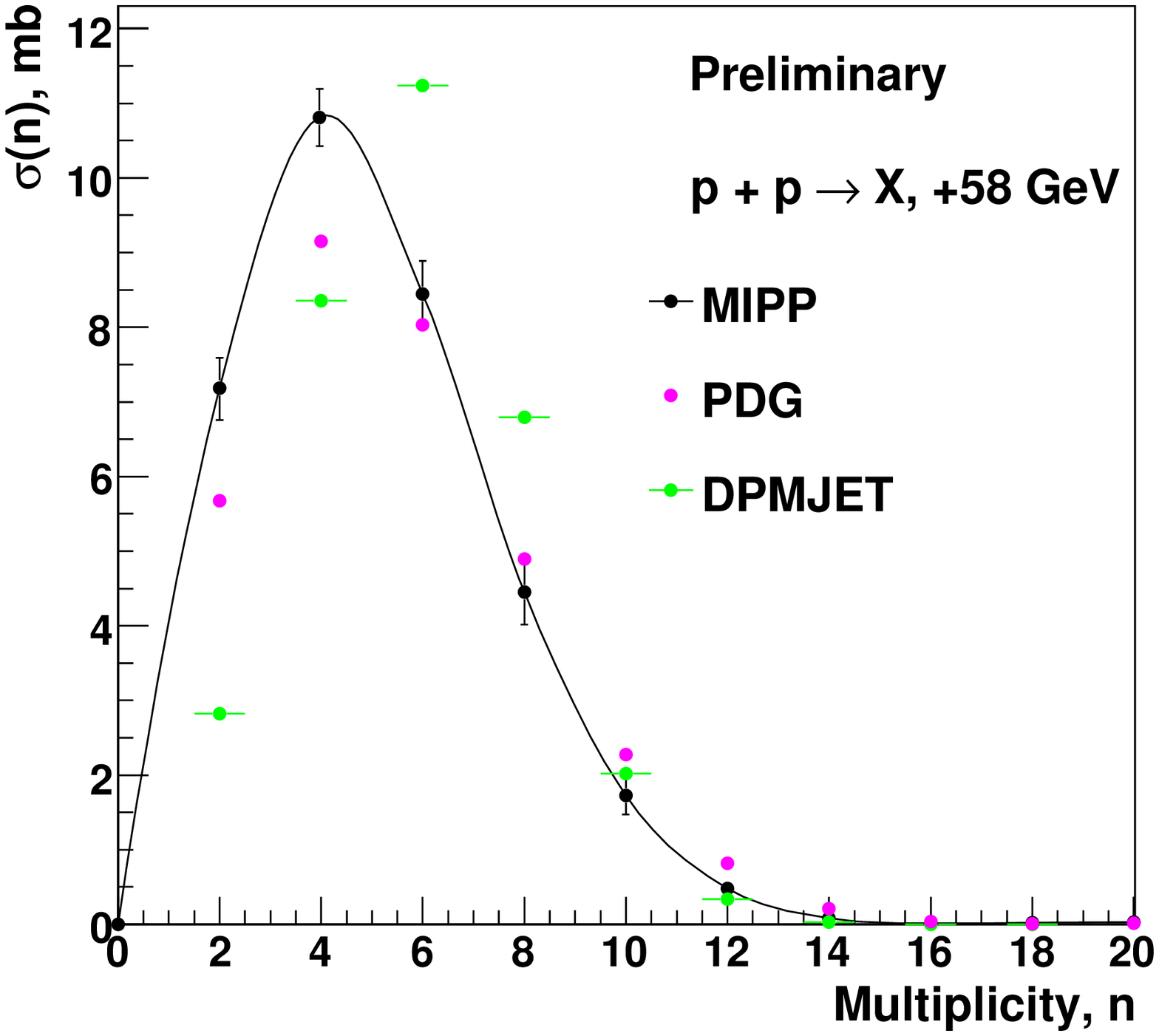}
}
\subfigure[]{
\includegraphics[width=2.5in,height=2.3in]{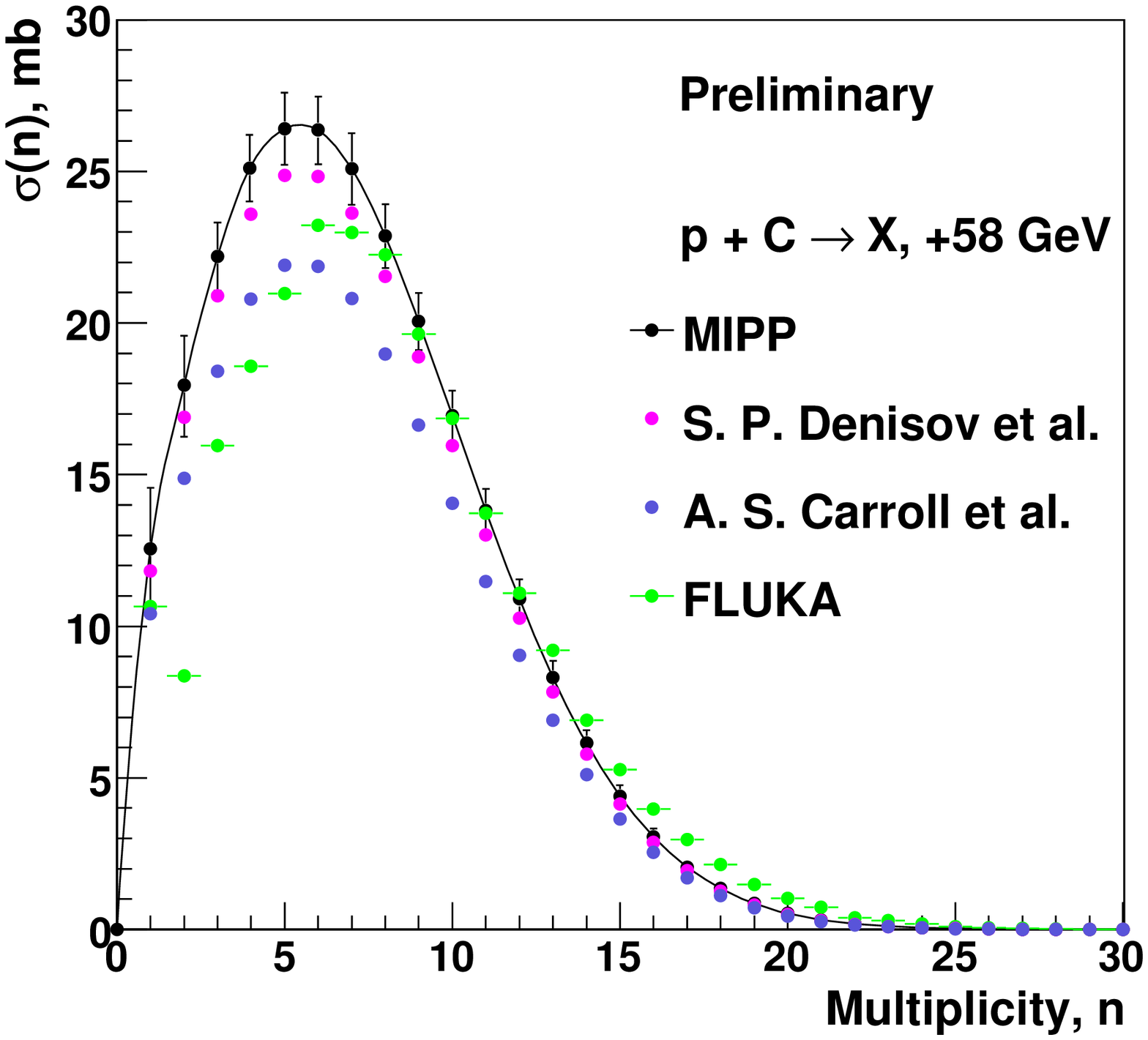}
}
\caption{Comparison of the MIPP data with the MC predictions and the previous measurements for 58 GeV/c p-H (a) and p-C (b) interactions.}
\end{center}
\end{figure*}
\section{Charged particle production cross sections}
Data and MC charged particle production cross sections have also been measured in bins of momentum, p$_{T}^{2}$ and x$_{F}$ for 58 and 120 GeV/c p-C interactions, and compared. For this, all charged particles are selected from the primary interaction. \\
Figures 5a, 5b and 5c show the comparison of the data and MC production cross sections for 58 GeV/c p-C interactions in bins of momentum, p$_{T}^{2}$ and x$_{F}$ respectively. The average production cross sections are calculated by summing over all the bins for both the data and MC for 58 and 120 GeV/c p-C interactions, and shown in Table 3. The MC is $\sim$12\% lower than the data at 58 GeV and $\sim$11\% lower than the data at 120 GeV. The average production cross section should be equal to average multiplicity times the average inelastic cross section. The average multiplicity is $\sim$7 and $\sim$8.5 in case of 58 and 120 GeV/c p-C interactions respectively. If we multiply it by the average inelastic cross section for both the data and MC, we get the numbers very close to the numbers shown in Table 3. \\
A drop of $\sim$8\% is observed in the cross section when it is binned in x$_{F}$. The x$_{F}$ of a particle is dependent on its mass and mass of the pion has been used for all the charged particles assuming that all of them are pions. It can be the reason for the observed drop in the cross section binned in x$_{F}$.
\begin{figure*}[ht]
\begin{center}
\subfigure[]{
\includegraphics[width=1.8in,height=1.8in]{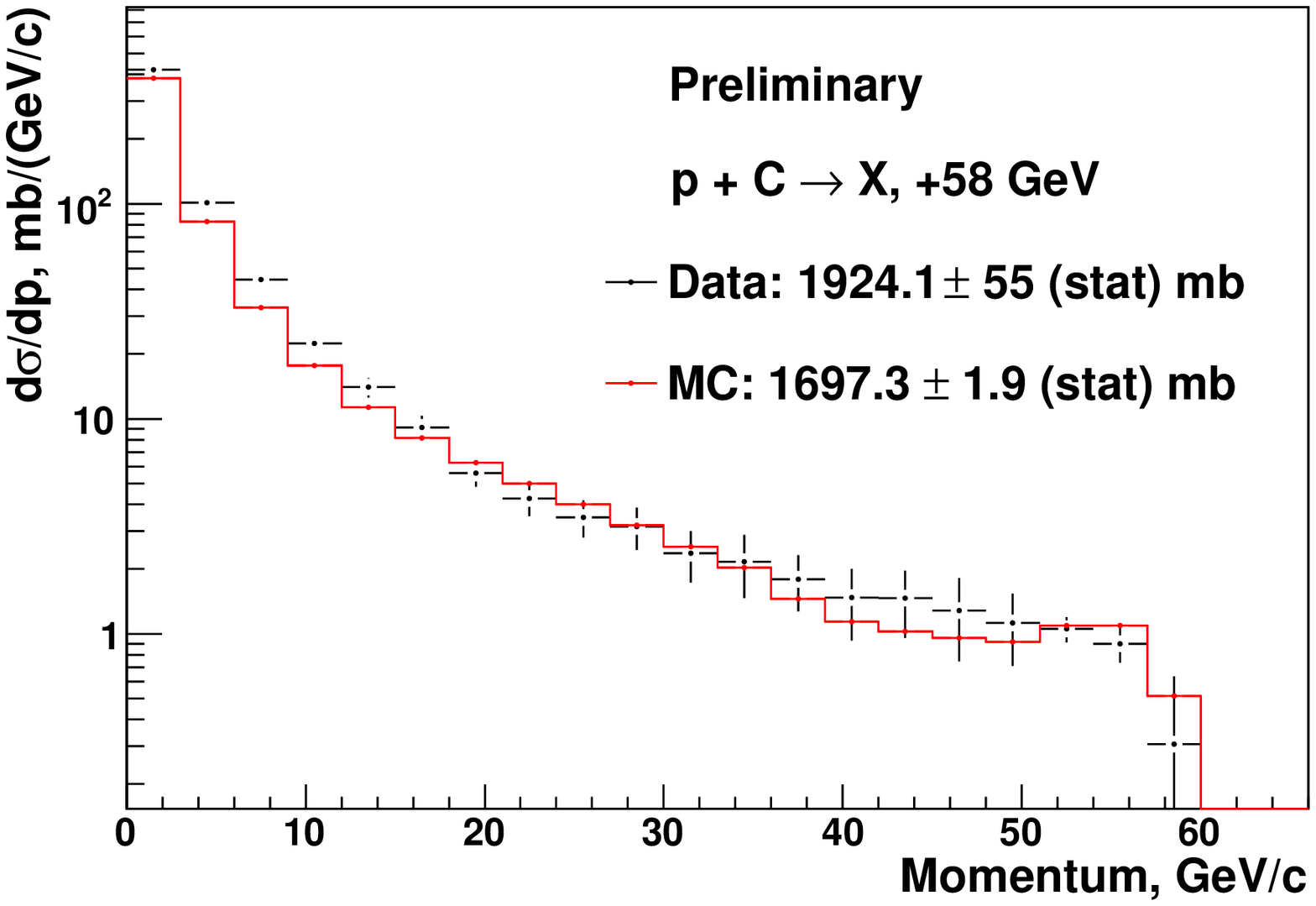}
}
\subfigure[]{
\includegraphics[width=1.8in,height=1.8in]{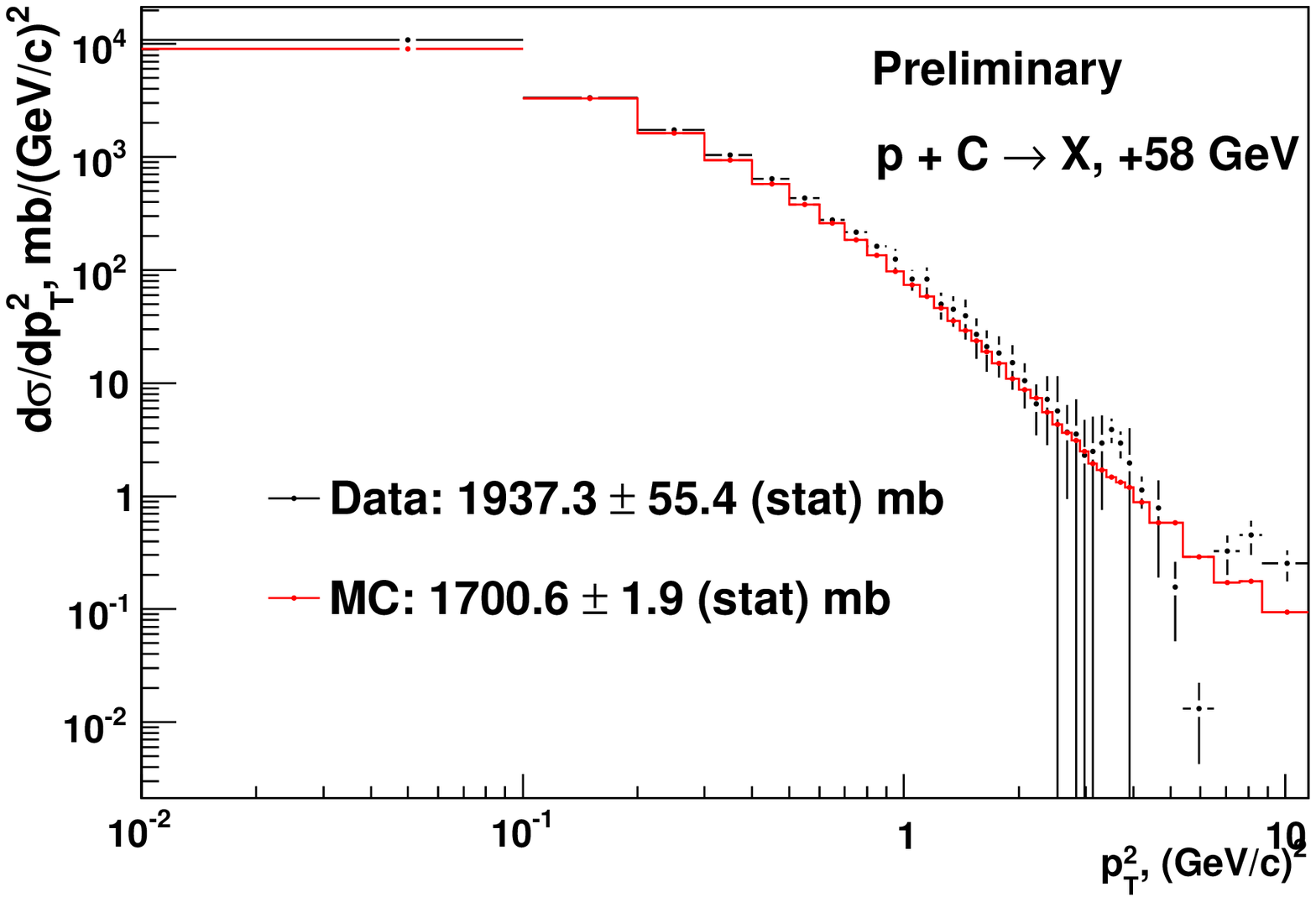}
}
\subfigure[]{
\includegraphics[width=1.8in,height=1.8in]{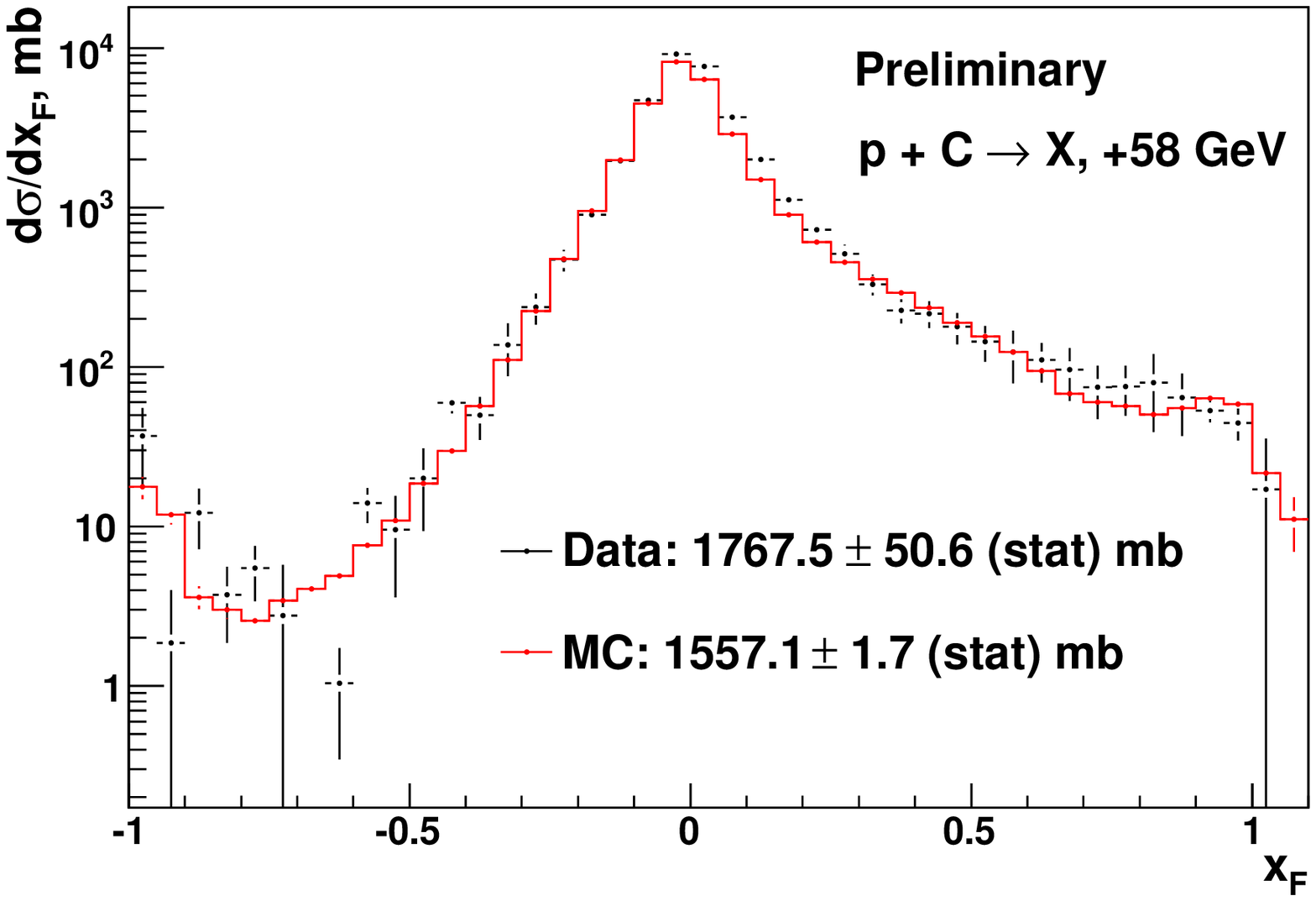}
}
\caption{Comparison of the data and MC charged particle production cross sections in bins of momentum (a), p$_{T}^{2}$ (b) and x$_{F}$ (c) for 58 GeV/c p-C interactions.}
\end{center}
\end{figure*}
\section{Particle identification}
Four hypotheses are considered for particle identification (PID) --e/$\pi$/K/p (denoted by H). TPC, ToF and RICH detectors are used and PID quantities dE/dx, time of flight and ring radii are measured. Maximum likelihood technique has been used to determine the spectra of each particle type in data. A weight is formed by using the likelihood of a PID quantity for a particular hypothesis and the sum of the likelihoods of that quantity for all the hypotheses and is calculated as:
\begin{center}
Weight = $\frac{Likelihood}{\Sigma_{H}Likelihoods}$
\end{center}
A global weight is formed using the total likelihood i.e. product of all the detector likelihoods. This is determined for each track and used to weight the track for each hypothesis. Each particle enters all hypothesis dependent plots with its hypothesis dependent weight. The aim is to determine the momentum spectrum for each particle type. \\
Figure 6 shows the Bethe Bloche dE/dx curves superimposed on the TPC data. The data is described well by the predictions. The global weighted mass squared distributions for the pion and kaon hypotheses from the ToF and RICH detectors using 120 GeV/c p-C data are shown in Figure 7. The means of these distributions are at exactly the masses of the pion and kaon. Small contamination from other particles can be seen in the kaon m$^{2}$ distribution from the RICH. We are working on the algorithm to remove this contamination. Figure 8 shows the momentum spectra for the negatively and positively charged pions and kaons from 120 GeV/c p-C data. The contamination can be seen at $\sim$120 GeV/c in the momentum spectrum for K$^{+}$ particles.\\
\begin{table}[t]
\small
\begin{center}
\scalebox{0.9}{
\begin{tabular}{l|cc}  
Energy (GeV) & Data (mb) & MC (mb)  \\ \hline
 58 & 1924.1$\pm55\rm(stat)^{+89.5}_{-94.2}$(syst)  &  1697.3$\pm1.9\rm(stat)\pm51.1$(syst) \\
 120 & 2224.5$\pm20.4\rm(stat)^{+92.5}_{-97.1}$(syst)  &  1986.5$\pm1.7\rm(stat)\pm51.5$(syst) \\ \hline
\end{tabular}}
\caption{Comparison of the data and MC average charged particle production cross section for p-C interactions at 58 and 120 GeV.}
\end{center}
\end{table}
\begin{figure}
\centering
\includegraphics[width=8cm,height=6cm]{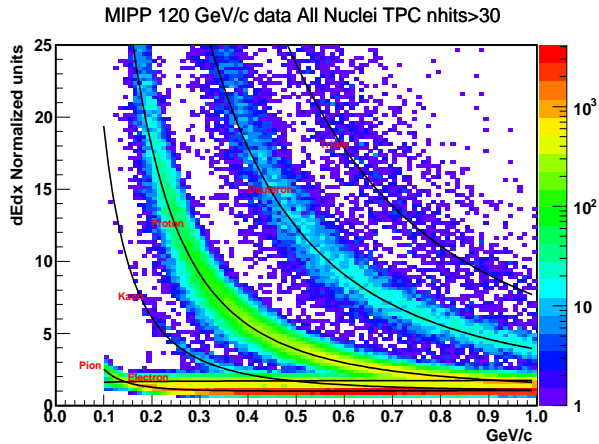}
\caption{Comparison of TPC data with the dE/dx predictions.}
\end{figure}
\begin{figure}
\centering
\includegraphics[width=4in,height=3in]{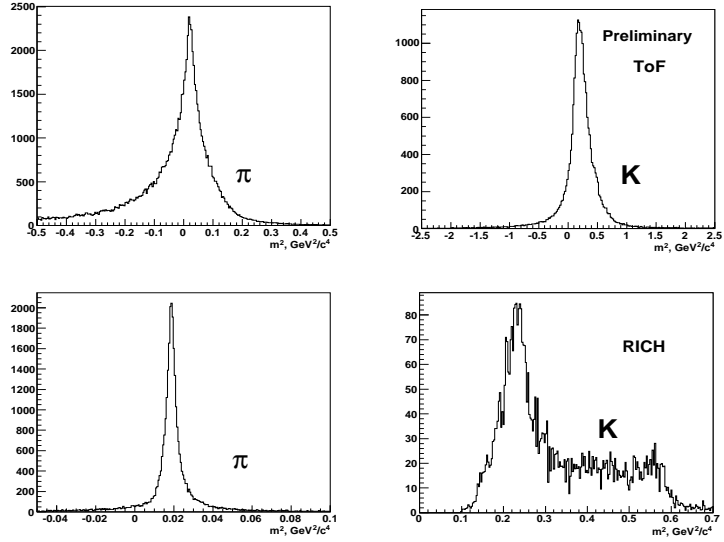}
\caption{Global weighted m$^{2}$ distributions for pions and kaons from the ToF (top) and RICH (bottom) detectors using 120 GeV/c p-C data.}
\end{figure}
\begin{figure}
\centering
\includegraphics[width=4in,height=3in]{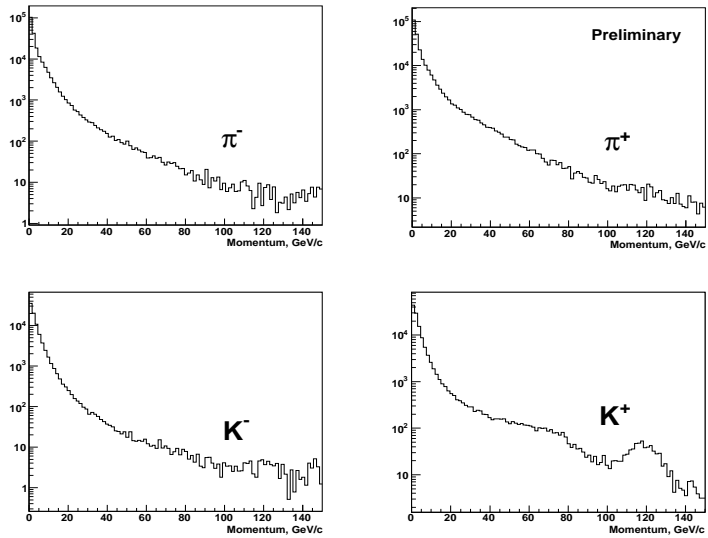}
\caption{Momentum spectra for $\pi^{\pm}$ (top) and K$^{\pm}$ (bottom) particles from 120 GeV/c p-C interactions.}
\end{figure}
We have to work out the inclusive pion and kaon production cross sections as a function of momentum for both the data and MC, and compare them. 

\end{document}

%% file: econfmacros.tex
\def\beq{\begin{equation}}
\def\eeq#1{\label{#1}\end{equation}}
\def\eeqn{\end{equation}}

\def\beqa{\begin{eqnarray}}
\def\eeqa#1{\label{#1}\end{eqnarray}}
\def\eeqan{\end{eqnarray}}

\let\bar=\overbar

\def\Dslash{\not{\hbox{\kern-4pt $D$}}}
\def\dslash{\not{\hbox{\kern-2pt $\del$}}}

\def\msb{{\bar{\ssstyle M \kern -1pt S}}}

